\documentclass[a4paper,11pt]{article}
\usepackage{pos}
\usepackage{bbm}

\numberwithin{equation}{section}
\newtheorem*{mydef}{Definition}
\newcommand{\M}{\mathcal{M}}
\newcommand{\R}{\mathbb{R}}
\newcommand{\E}{\mathbb{E}}
\newcommand{\p}{\partial}
\newcommand{\rd}{\mathrm{d}}
\newcommand{\ri}{\mathrm{i}}

\title{Quantum Theory, Gravity and Higher Order Geometry}

\author[a]{Folkert Kuipers}

\affiliation[a]{Arnold Sommerfeld Center for Theoretical Physics, Ludwig-Maximilians-Universit\"at M\"unchen,\\ Theresienstra{\ss}e 37, 80333 M\"unchen, Germany}

\emailAdd{f.kuipers@lmu.de}

\abstract{The fact that quantum theory is non-differentiable, while general relativity is built on the assumption of differentiability sources an incompatibility between quantum theory and gravity. Higher order geometry addresses this issue directly by extending differential geometry, such that it can be applied to theories that are non-differentiable, but have a certain degree of H\"older regularity. As this includes the path integral formulation of quantum theory, it provides a natural mathematical framework for describing the interplay between gravity and quantum theory. In this article, we review the motivation for and the basic features of this framework and point towards future developments.}

\FullConference{Proceedings of the Corfu Summer Institute 2024 "School and Workshops on Elementary Particle Physics and Gravity" (CORFU2024)\
12 - 26 May, and 25 August - 27 September, 2024\\
Corfu, Greece\\}

\tableofcontents

\renewcommand{\logo}{\relax}

\begin{document}

\maketitle

\section{Introduction}\label{sec:intro}
In his seminal paper from 1948, Feynman \cite{Feynman:1948ur}, inspired by the works of Dirac, introduced a new formulation of quantum mechanics that is equivalent but independent of the earlier formulations provided by Schr\"odinger and Heisenberg. 
According to this path integral formulation, a quantum particle follows a well-defined trajectory trough spacetime. However, due to a fundamental randomness, it is impossible to predict this trajectory with certainty. Instead, one can only calculate the probability that a certain trajectory is followed. Thus, in order to evaluate the dynamics of the particle, one must properly average over all the possible trajectories.
\par 

Since its introduction, the path integral formulation has been applied very successfully as a computational tool for obtaining the correlation functions in quantum field theories in the absence of gravity. In the presence of gravity, the path integral is at odds with general covariance, leading to various gravitational anomalies in the quantum theory, which necessitates the formulation of manifestly covariant quantum theories \cite{DeWitt:1957at,DeWitt:1967ub,ChapelHill,DeWitt:2003pm}. 
This incompatibility between gravity and quantum theory is inherently related to the fractal nature of the paths that appear in the path integral \cite{Feynman:1948ur,Abbott:1979bh,Koch:2014cma}, which will serve as our main motivation for introducing higher order geometry.
\par  

In order to discuss this issue in more detail, we review some elementary aspects of the path integral in the worldline theory of a point particle on curved space(time), which can be interpreted as a 1-dimensional quantum field theory.
The guiding philosophy of the path integral formulation is to replace this deterministic worldline by an uncountable collection of worldlines that can all occur with a certain probability:
\begin{equation}
	X : \mathcal{T} \rightarrow \M 
	\qquad \overset{\rm Quantitization}{\longrightarrow} \qquad 
	X : \mathcal{T} \times \Omega \rightarrow \M \, ,
\end{equation}
where $\M$ is some Riemannian or Lorentzian manifold, $\mathcal{T}=[0,T]\subset\R$ is a time interval\footnote{In a non-relativistic theory $t\in\mathcal{T}$ is the physical time parameter, while in a non relativistic theory $\lambda\in\mathcal{T}$ labels an arbitrary affine parameter. For a massive particle, this parameter can be fixed to be the proper time $\lambda=\tau$  by gauge fixing the reparameterization invariance. Similarly, for tachyons this parameter can be fixed to be the proper distance $\lambda=s$.} and $\Omega$ is a sample space. The elements $\omega\in\Omega$ that label the paths are drawn according to some probability measure $\mathbb{P}:\Sigma(\Omega)\rightarrow[0,1]$, where $\Sigma(\Omega)$ is some sigma algebra over $\Omega$. The transition amplitude for a state localized at $X(\tau_0)=x_0$ to a state localized at $X(\tau_f)=x_f$ can now be expressed as
\begin{equation}
	\langle x_T, T \, | \, x_0 , 0 \rangle = \int_{X(0)=x_0}^{X(T)=x_T} e^{\ri \, S(X)} DX \, ,
\end{equation}
where $S$ is the action, and $e^{\ri \, S(X)}$ is conjectured to be the probability density for the trajectory $X$ to occur, such that\footnote{Here, we assume the existence of the probability measure $\mathbb{P}$. The existence of this probability measure poses an infamous problem to the path integral formulation. For the Euclidean theory, it exists, as $\mu$ can be related to the Wiener measure. However, in the Lorentzian theory it is not yet clear whether there exists a well-defined probability measure $\mathbb{P}$ that provides a probabilistic interpretation to the path integral, cf. e.g. \cite{Albeverio} for more detail.}
\begin{equation}\label{eq:PImeasure}
	\rd(\mathbb{P}\circ X^{-1}) = \rd\mu(X) = \rho(X) \, DX \sim \exp\left[\ri \, \int_\mathcal{T} L(X,\dot{X},t) \, dt \right] \, DX \, .
\end{equation}
The correlation functions of the theory can then be obtained using the partition function
\begin{equation}
	Z[J] = \int e^{ \ri \left[ S(X) + \langle J, X\rangle \right] } \, DX 
\end{equation}
or using the characteristic functional\footnote{If the probability density $\rho(X)$ exists, one can also define a moment generating functional $M_X(J)= \phi_X(-\ri \, J)$.}
\begin{equation}
	\phi_X(J) 
	= \E\left[ e^{\ri \, \langle J, X\rangle }\right] 
	= \int_\Omega e^{\ri \,\langle J, X\rangle } \, \rd\mathbb{P}(\omega)
	= \int_{L^2(\Omega)} e^{\ri \,\langle J, X\rangle } \, \rd\mu(X) \, ,
\end{equation}
which is already normalized such that 
\begin{equation}\label{eq:CharPart}
	\phi_X(J) = \frac{Z(J)}{Z(0)} \, .
\end{equation}

An important feature of the path integral is that the differentiable paths have $0$ measure in the path integral. This was already pointed out by Feynman, who used this as a key property for establishing the equivalence of the path integral with the canonical formulation of quantum theory \cite{Feynman:1948ur}. Inspired by this work, Kac \cite{Kac:1949} proved the Feynman-Kac theorem which provides a mathematical relation between Euclidean quantum theory and the Wiener process, showing that the paths in the Euclidean path integral are of the same type as the sample paths of a Brownian motion, as already suggested by Feynman \cite{Feynman:1948ur}. Hence, the quantum paths have the same H\"older regularity as Brownian motion, for which we recall the definition:
\begin{mydef} {\rm (H\"older Continuity)}
	A trajectory $X:\mathcal{T}\rightarrow\R^n$ is $\alpha$-H\"older continuous if there exists $C\geq0$ such that $||X(t)-X(s)||\leq C \, |t-s|^\alpha$ for all $s,t\in\mathcal{T}$.
\end{mydef}
\noindent Then, we have the following properties\footnote{Note that the properties are not completely independent as the upper bound on the Hausdorff dimension is fixed by the H\"older regularity as $\dim_H \{X\} \leq \alpha_{\sup}^{-1}$.} \cite{Morters:2010}:
\begin{enumerate}
	\item \textbf{Roughness}:\label{rough} The paths in the path integral are, with probability $1$, H\"older continuous everywhere for the H\"older parameter $\alpha<1/2$, but nowhere for $\alpha>1/2$.
	\item \textbf{Worldsheet}:\label{worldsheet} The Hausdorff dimension of the set of paths is $\dim_H\{X(t)\, | \, t\in\mathcal{T}\} = \min\{2,n\}$, where $n$ is the space(time) dimension.
\end{enumerate}
In contrast, classical general relativity implicitly assumes that any theory defined on the spacetime manifold follows trajectories that are at least twice continuously differentiable. Since this assumption is  violated by the paths appearing in the path integral formulation, one finds an incompatibility between the two theories. Moreover, differential geometry, which forms the mathematical foundation of general relativity, requires trajectories to be at least once continuously differentiable. 
\par 

Any theory that couples quantum theory to gravity, without abandoning the path integral formalism, will have to address this incompatibility. For this, one has roughly three options:
\begin{itemize}
	\item One may regularize the short scale behavior in the path integral, such that the paths become once (or even twice) continuously differentiable. In this case, methods from differential geometry can still be applied.
	\item One may modify general relativity and differential geometry, such that it becomes compatible with the rough paths encountered in the path integral.
	\item One may use that the path integral can be obtained as a continuum limit of a universality class of discretized paths parameterized along a discrete time parameter, which provides a natural regularization of the short scale behavior of the path integral.
\end{itemize}
At large scales the differences between the approaches should all be perturbatively suppressed by some cutoff scale, and one can apply effective field theory methods to describe the interplay between quantum theory and gravity. However, at short scales the approaches will generically lead to different predictions.
\par 
	
The first and third option are most commonly followed in the study of the interplay between gravity and quantum theory. The first option has the advantage that one does not require a geometrical framework beyond standard (pseudo-)Riemannian geometry.\footnote{The introduction of such frameworks is not excluded, but there is no strict necessity to do so.} However, it does require to conjecture a new framework that modifies quantum theory at short distance scales, for example by changing the fundamental degrees of freedom. The third scenario, on the other hand, introduces some discreteness of spacetime. As both quantum theory, assuming the standard quantization prescriptions,\footnote{If one allows for generalizations of the quantization prescription, quantum theory does not require the assumption of spacetime continuity. For example, if one considers a quantization involving some spacetime non-commutavity on top of the usual canonical quantization prescription, one may find discrete eigenspectra for the position operators. Such modifications are typically accompanied by the introduction of a curved momentum space, cf. e.g. \cite{Franchino-Vinas:2023rcc}.} and general relativity are built on the assumption of spacetime continuity, this introduces a modification of both gravity and quantum theory at small length scales.
\par

In this work, we will focus on the second approach. The advantage of this approach is that it does not necessarily require a modification of the small distance behavior of quantum theory. However, it does require a generalization of differential geometry that is compatible with continuous, but non-differentiable theories. This modification is provided by higher order geometry \cite{Bies:2023zvs}, which interpolates between differential and discrete geometry. Generically speaking, a theory with maximal H\"older parameter $\alpha_{\sup} =1/k$ requires the study of $k$-th order geometry. Hence, for differentiable trajectories one only requires $1$st-order geometry, i.e. differential geometry, whereas $\infty$-order geometry can be applied to certain classes of discrete trajectories. Since the path integral satisfies $\alpha_{\sup} =1/2$, we will focus in this work on second order geometry \cite{Schwartz:1984,Meyer:1981,Emery:1989,Huang:2022,Kuipers:2024gfp}.

\section{Rough paths}\label{sec:Rough}
An immediate consequence of the roughness property is the divergence of the limit
\begin{equation}
	\lim_{\epsilon\rightarrow 0} \frac{X(t+\epsilon) - X(t)}{\epsilon} 
	\sim \lim_{\epsilon\rightarrow 0} \frac{1}{\sqrt{\epsilon}} 
	= \infty \, ,
\end{equation}
which would usually provide the velocity along the path. Let us now make the assumption that this divergence can be isolated, such that we can write the increments for the left and right limit as
\begin{alignat}{2}
	\rd_+ X(t) &= X(t+dt) - X(t) &&= b_+ \, dt^{1/2} + v_+ \, dt + o(dt) \, ,\nonumber\\
	\rd_- X(t) &= X(t) - X(t-dt) &&= b_- \, dt^{1/2} + v_- \, dt + o(dt) \, .\label{eq:ItoEq}
\end{alignat}
Alternatively, by performing a change of basis, we may write these objects as a velocity increment, constructed as the average of the left and right limit, and a acceleration increment, constructed as the difference of the two limits:
\begin{alignat}{2}
	\rd_0 X(t) &= X(t+dt/2) - X(t-dt/2) &&= \frac{b_+ + b_-}{\sqrt{2}} \, dt^{1/2} + v_\circ \, dt + o(dt) \, ,\nonumber\\
	\frac{1}{2} \rd_0^2 X(t) &= \frac{X(t+dt) - 2\, X(t) + X(t-dt)}{2} &&= \frac{ b_+ - b_-}{2} \, dt^{1/2} + v_\perp \, dt + o(dt) \, , \label{eq:StratEq}
\end{alignat}
where we introduced the \textit{current velocity} and \textit{osmotic velocity} given by
\begin{equation}\label{eq:vcircvperp}
	v_\circ = \frac{v_+ + v_-}{2} \qquad {\rm and} \qquad v_\perp = \frac{v_+ - v_-}{2} \, .
\end{equation}
Let us also note that the square of the increments does not vanish at order $dt$, and is given by
\begin{align}\label{eq:QVarMatrix}
	\begin{pmatrix}
		\rd_+ X^i(t) \, \rd_+ X^j(t') & \rd_+ X^i(t) \, \rd_- X^j(t') \\
		\rd_- X^i(t) \, \rd_+ X^j(t') & \rd_- X^i(t) \, \rd_- X^j(t')
	\end{pmatrix}
	&=
	\begin{pmatrix}
		b_+^i(t) \, b_+^j(t') & b_+^i(t) \, b_-^j(t') \\
		b_-^i(t) \, b_+^j(t') & b_-^i(t) \, b_-^j(t')
	\end{pmatrix}
	dt + o(dt) \, .
\end{align}

\subsection{Relation to stochastic calculus}\label{sec:StochCalc}
Due to the appearance of the $dt^{1/2}$ terms, the equations \eqref{eq:ItoEq} are not well-defined within ordinary calculus. However, they can be made well-defined within fractional calculus, where they appear as the fractional derivative $\rd^{1/2} X/\rd t^{1/2}$, or in the framework of stochastic calculus. In the latter, these equations represent stochastic differential equations and their solutions are stochastic processes, i.e. a set of paths $\{X(\omega,t)\, |\, t\in\mathcal{T}, \omega\in\Omega\}$ together with a probability measure $\mathbb{P}$ that determines the probability for each path $X(\omega):\mathcal{T}\rightarrow \M$.
\par 

In stochastic calculus, one writes
\begin{equation}
	\rd_\pm M(t) = b_\pm(t)\, dt^{1/2} \, ,
\end{equation}
and requires that it satisfies the martingale property:
\begin{equation}
	\E\left[ \rd_\pm M(t) \, | \, X(t)\right] = 0 \, ,
\end{equation}
such that the drift velocity is given by 
\begin{equation}
	\lim_{dt\rightarrow 0} \E\left[ \frac{\rd_\pm X^i(t)}{\rd t} \, \Big| \, X(t)=x\right] = v_\pm^i(x,t) \, .
\end{equation}
In that context, the square of the increments \eqref{eq:QVarMatrix} is called the quadratic variation and its derivative defines a second order velocity field given by
\begin{align}
	\lim_{dt\rightarrow0} \E\left[ \frac{\rd_\pm X^i(t) \, \rd_\pm X^j(t')}{dt} \, \Big| \, X(t)=x \right]
	&= v_{\pm\pm}^{ij}(x,t,t') \, ,\\
	\lim_{dt\rightarrow0} \E\left[ \frac{\rd_\pm X^i(t) \, \rd_\mp X^j(t')}{dt} \, \Big| \, X(t)=x \right]
	&= v_{\pm\mp}^{ij}(x,t,t') \, .
\end{align}
We will be interested in the case in which this field is given by
\begin{equation}\label{eq:SecVel}
	\begin{pmatrix}
		v_{++}^{ij}(x,t,t') & v_{+-}^{ij}(x,t,t') \\
		v_{-+}^{ij}(x,t,t') & v_{--}^{ij}(x,t,t')
	\end{pmatrix}
	= 
	\frac{\hbar}{m} \, g^{ij}(x) \, \delta(t-t') 
	\begin{pmatrix}
		\alpha_{++} & \alpha_{+-} \\
		\alpha_{-+} & \alpha_{--}
	\end{pmatrix} ,
\end{equation}
where the $\delta$-function imposes the theory to be local in time. Moreover, $g^{ij}$ is the inverse metric and $m$ is the mass of the particle following the trajectory.\footnote{Assuming a non-relativistic theory. For a relativistic theory, one replaces $t\rightarrow\lambda$ and $m\rightarrow \varepsilon^{-1}(\lambda)$, where $\lambda$ is an arbitrary affine parameter and $\varepsilon$ is an auxiliary field, cf. section \ref{sec:RLT}.} Since $X$ is real valued, we may impose $\alpha_{\pm\pm},\alpha_{\pm\mp}\in\R$. Moreover, since this defines a covariance matrix, whose eigenvalues must be positive, one obtains the constraints\footnote{Assuming that the manifold is Riemannian, such that the metric has a positive definite signature. On a pseudo-Riemannian manifold, one must introduce two parameters: $\alpha_t$ associated to the temporal directions (negative eigenvalues of the metric) and $\alpha_s$ associated to the spatial directions (positive eigenvalues of the metric). Then, both $\alpha_s,\alpha_t$ satisfy the first condition, and the second condition is satisfied for $\alpha_s$. However, for $\alpha_t$ the second condition is modified to $\alpha_{++}+\alpha_{++}\leq-\sqrt{(\alpha_{++}-\alpha_{--})+4\,\alpha_{+-}\alpha_{-+}}$, cf. e.g. \cite{Dohrn:1985iu,Kuipers:2023pzm}.}
\begin{align}
	\left(\frac{\alpha_{++} - \alpha_{--}}{2}\right)^2 \, &\geq \, - \alpha_{+-} \alpha_{-+} \, ,
	\label{ConstrAlphaI}\\
	\frac{\alpha_{++} + \alpha_{--}}{2} \, &\geq \, \sqrt{ \left(\frac{\alpha_{++} - \alpha_{--}}{2} \right)^2 + \alpha_{+-} \alpha_{-+}} \, .
\end{align}
Let us make a few remarks
\begin{itemize}
	\item If $\alpha_{+-}=\alpha_{-+}=0$, the increments $\rd_+X$ and $\rd_-X$ are stochastically independent. Due to locality of the second order velocity field, this implies that all increments are independent, which in turn implies that the process satisfies the Markov property. Therefore, if $\alpha_{\pm\mp}=0$, the process is Markovian.
	\item If $\alpha_{+-}=\alpha_{-+}$, the increments $\rd_+X$ and $\rd_-X$ commute, such that  $[b_+^i,b_-^j]=0$. In this case, the constraint \eqref{ConstrAlphaI} is trivially satisfied. Therefore, this constraint puts a bound on the non-commutativity of these increments.
	\item The commutation relations $[b_+^i,b_+^j]=0$ and $[b_-^i,b_-^j]=0$ imply a commutation of the increment $\rd_+X^i$ with $\rd_+X^j$ and of $\rd_-X^i$ with $\rd_-X^j$, such that the second order velocity field $v_2$ is symmetric. Here, this commutation is assumed, since we have fixed $v_2^{ij}\propto g^{ij}$.
\end{itemize}
Let us also provide a few examples of processes described by these equations:
\begin{itemize}
	\item If $\alpha_{++}=\alpha>0$ and $\alpha_{\pm\mp}=\alpha_{--}=0$, the process is a standard Wiener process with diffusion coefficient $\frac{\alpha\, \hbar}{m}$.
	\item If $\alpha_{--}=\alpha>0$ and $\alpha_{\pm\mp}=\alpha_{++}=0$, the process is a standard Wiener process with diffusion coefficient $\frac{\alpha\, \hbar}{m}$ that evolves backward in time.
	\item If $\alpha_{\pm\pm}=\alpha>0$ and $\alpha_{\pm\mp}=0$, the process is a two-sided Wiener process \cite{Nelson:1967,Nelson:1985} with diffusion coefficient $\frac{\alpha\, \hbar}{m}$ .
	\item If $\alpha_{\pm\pm}\geq0$ and $\alpha_{+-}=\alpha_{-+}=\pm\sqrt{\alpha_{++}\alpha_{--}}$, one obtains the processes described in Refs.~\cite{Kuipers:2023pzm,Kuipers:2023ibv}.
\end{itemize}

Finally, we remark that stochastic analysis has a relation to canonical quantization, in the sense that some stochastic processes $M$ can be reinterpreted as operators acting on a Fock space. In particular, let us set the dimension\footnote{Cf. \cite{Biane:2010} for the generalization to $n>1$.} $n=1$, and consider the Fock space $\mathcal{F}(L^2(\R_+))$ with states $|k \rangle$, then one has the following results \cite{Biane:2010}.
\begin{itemize}
	\item If $M(t)$ is a standard Wiener process, one can decompose $M=a_-+a_+$, where $a_\pm$ are creation and annihilation operators. Thus, $a_-|k\rangle \propto |k-1\rangle$, $a_+|k\rangle \propto |k+1\rangle$, such that they satisfy the canonical commutation relations $[a_-,a_+]= \mathbbm{1}$.
	\item If $N(t)$ is a standard Poisson process, the process $M=N- t$ is a compensated Poisson process, which can be decomposed as $M=a_-+ a_0+a_+$, where $a_\pm,a_0$ are creation, annihilation and preservation operators. Thus, $a_-|k\rangle \propto |k-1\rangle$, $a_0|k\rangle \propto |k\rangle$ and $a_+|k\rangle \propto |k+1\rangle$, such that they satisfy the commutation relations $[a_-,a_0]= a_- $, $[a_0,a_+]= a_+ $ and $[a_-,a_+]= a_0$.
\end{itemize}
This suggests that the objects $\int b_\pm \, dt^{1/2}$ can be reinterpreted in terms of creation and annihilation operators and that solutions $X$ of the differential equation \eqref{eq:StratEq} can be reinterpreted as position operators in the Heisenberg picture.

\section{Second order geometry}\label{sec:2ndorder}
Using the differentials \eqref{eq:ItoEq}, one finds that a total derivative along a scalar field $\phi$ is given by
\begin{align}\label{eq:TotDer}
	\rd_\pm \phi
	&= \p_\mu \phi(x) \, \rd_\pm x^\mu \pm \frac{1}{2} \p_\nu\p_\mu \phi(x) \, \rd_\pm x^\mu \rd_\pm x^\nu + o(dt) \nonumber\\
	&= b_\pm^\mu \p_\mu \phi(x) \, dt^{1/2} + \left(  v_\pm^\mu \p_\mu \phi(x) \pm \frac{1}{2} \, v_{\pm\pm}^{\mu\nu} \p_\nu\p_\mu \phi(x) \right) dt + o(dt) \, ,
\end{align}
which implies modifications of the usual first order product rule and chain rule:
\begin{align}
	\rd_\pm( \phi \, \psi) &= \phi \, \rd_\pm \psi + \psi \, \rd_\pm \phi \pm \rd_\pm\phi \, \rd_\pm \psi \, ,\label{Leibniz}\\
	\rd_\pm( f \circ \phi) &= (f'\circ\phi) \, \rd_\pm\phi \pm \frac{1}{2} (f''\circ \phi) \, \rd_\pm\phi \, \rd_\pm\phi \, .
\end{align}
Second order geometry \cite{Schwartz:1984,Meyer:1981,Emery:1989} is a geometrical framework based on such a modification of the total derivative operator. Using eq.~\eqref{eq:TotDer}, one finds that this involves the following generalization of the notion of a vector field
\begin{equation}
	v \in T\M  \quad {\rm s.t.} \quad v = v^\mu\p_\mu \qquad \rightarrow \qquad  v_\pm \in T_2\M  \quad {\rm s.t.} \quad v_\pm = v_\pm^\mu\p_\mu \pm \frac{1}{2} v_{\pm\pm}^{\mu\nu} \p_\nu\p_\mu \, .
\end{equation}
Usually \cite{Schwartz:1984,Meyer:1981,Emery:1989}, second order  geometry focuses solely on the right limits $v_+^{\mu},v_{++}^{\mu\nu}$. In this case, the second order tangent space is given by
\begin{equation}
	T_{2,x}\M \cong (T_x\M) \oplus ( T_x\M \otimes T_x\M)
\end{equation}
with corresponding tangent bundle 
\begin{equation}
	T_2\M = \bigsqcup_{x\in\M} T_{2,x}\M \, ,
\end{equation}
whose structure group is the It\^o group \cite{Huang:2022}. This is the semi-direct product 
\begin{equation}
	G_n^I = {\rm GL}(n,\R)\ltimes{\rm Lin}(\R^n\otimes\R^n,\R^n)
\end{equation}
with binary operation, for all $g,g'\in{\rm GL}(n,\R)$ and $\kappa,\kappa'\in{\rm Lin}(\R^n\otimes\R^n,\R^n)$,
\begin{equation}
	(g',\kappa') \, (g,\kappa) = (g' \, g, \, g'\circ\kappa + \kappa'\circ(g\otimes g)) \, .
\end{equation}
This group has a left action on the tangent spaces $T_{2,x}\M$, for all $(g,\kappa)\in G^n_I$ and $(v,v_2)\in T_{2,x}\M$  given by
\begin{equation}\label{GroupAction}
	(g,\kappa) \, (v,v_2) = (g \, v + \kappa \, v_2, \, (g\otimes g) \, v_2) \, .
\end{equation}
\par 

In principle, a physical theory defined within second order geometry could depend on any of the velocity fields $v_{\pm}^\mu,v_{\pm\pm}^{\mu\nu},v_{\pm\mp}^{\mu\nu}$. Hence, in full generality, the second order tangent space will be given by
\begin{align}\label{eq:Tangentspace}
	T_{2,x}\M &\cong (T_x\M)_+ \oplus (T_x\M)_- \oplus ( T_x\M \otimes T_x\M)_{++} \oplus ( T_x\M \otimes T_x\M)_{--} \nonumber\\
	&\quad \oplus ( T_x\M \otimes T_x\M)_{+-} \oplus ( T_x\M \otimes T_x\M)_{-+}
\end{align}
In section \ref{sec:ToyModel}, we will consider a toy model for such a theory. There, we consider a theory depending on the velocity fields
\begin{equation}
	\begin{pmatrix} 
		\mathfrak{v} &
		\mathfrak{v}_2
	\end{pmatrix}
	= 
	\begin{pmatrix} 
		v_\circ &
		v_{2\circ}
	\end{pmatrix}
	+ \ri 
	\begin{pmatrix} 
		v_\perp &
		v_{2\perp}
	\end{pmatrix} \, ,
\end{equation}
where $v_\circ,v_\perp$ are defined in eq.~\eqref{eq:vcircvperp} and $v_{2\circ},v_{2\perp}$ are defined analogously by
\begin{equation}
	v_{2\circ} = \frac{v_{2+} + v_{2-}}{2} \qquad {\rm and} \qquad v_{2\perp} = \frac{v_{2+} - v_{2-}}{2} 
\end{equation}
with 
\begin{equation}
	v_{2\pm} = \pm v_{\pm\pm} \, .
\end{equation}
Hence, in the toy model discussed in section \ref{sec:ToyModel}, we consider the tangent space
\begin{equation}
	T_{2,x}\M \cong (T_x\M)^{\mathbb{C}} \oplus ( T_x\M \otimes T_x\M)^{\mathbb{C}} \, .
\end{equation}

\subsection{Covariance}\label{sec:Covariance}
An important feature of second order geometry is that the group action \eqref{GroupAction} mixes the first and second order sectors in the tangent bundle, such that the vectors do not transform covariantly under general coordinate transformations. Instead, one finds that a vector $(v^\mu,v_2^{\rho\sigma})$ transforms as 
\begin{equation}\label{eq:TransMat}
	\begin{pmatrix}
		v^\mu \\ v_2^{\rho\sigma}
	\end{pmatrix}
	\rightarrow
	\begin{pmatrix}
		\tilde{v}^\mu \\\tilde{v}_2^{\rho\sigma}
	\end{pmatrix}
	= 
	\begin{pmatrix}
		\frac{\p \tilde{x}^\mu}{\p x^\nu} & \frac{1}{2} \, \frac{\p^2 \tilde{x}^\mu}{\p x^\kappa \p x^\lambda}\\
		0 & \frac{\p \tilde{x}^\rho}{\p x^\kappa} \frac{\p \tilde{x}^\sigma}{\p x^\lambda}
	\end{pmatrix}
	\begin{pmatrix}
		v^\nu \\ v_2^{\kappa\lambda}
	\end{pmatrix}
\end{equation}
and a similar expression can be obtained for forms. This suggests to introduce a notion of covariant vectors $(\hat{v}^\mu,\hat{v}_2^{\mu\nu})$ and forms $(\hat{p}_\mu,\hat{p}_{\mu\nu})$ given by
\begin{alignat}{3}\label{eq:CovVec}
	\hat{v}^\mu &:= v^\mu + \frac{1}{2} \, \Gamma^\mu_{\rho\sigma} v_2^{\rho\sigma} \, ,
	\qquad \qquad && \hat{p}_\mu &&:=p_\mu \, , \nonumber\\
	\hat{v}_2^{\mu\nu} &:= v_2^{\mu\nu} \, ,
	\qquad \qquad && \hat{p}_{\mu\nu} &&:= p_{\mu\nu} - \frac{1}{2} \, \Gamma^\rho_{\mu\nu} p_\rho  \, ,
\end{alignat}
where $\Gamma$ are the Christoffel symbols.
For these covariant representations the transformation matrix \eqref{eq:TransMat} diagonalizes.
\par 

Note that, if the manifold is torsion-free, only the symmetric part of $v_2$ is relevant, such that one may restrict the tangent spaces to $T_x\M \otimes T_x\M\rightarrow{\rm Sym}(T_x\M \otimes T_x\M)$. The same restriction can also be obtained by imposing the commutation relations $[b_\pm,b_\pm]=0$, cf. section \ref{sec:StochCalc}.

\subsection{Metric}\label{sec:Metric}
As our intention is to describe a theory of gravity, we require a notion of a metric, leading to a (pseudo-)Riemannian version of second order geometry. We impose the usual requirement on the metric, such that it is a function $G:T_{2,x}\M\times T_{2,x}\M\rightarrow \R$ 
that is symmetric. Hence, we consider the following generalization
\begin{equation}
	g_{\mu\nu} \rightarrow 
	\begin{pmatrix}
		G_{\mu\nu} & G_{\mu|\kappa\lambda|} \\
		G_{|\rho\sigma|\nu} & G_{|\rho\sigma|\kappa\lambda|}
	\end{pmatrix}
\end{equation}
with the symmetry conditions
\begin{equation}
	G_{\mu\nu} = G_{\nu\mu}
	\, , \qquad
	G_{\mu|\rho\sigma|} = G_{|\rho\sigma|\mu}
	\, , \qquad
	G_{|\rho\sigma|\kappa\lambda|} = G_{|\kappa\lambda|\rho\sigma|} \, .
\end{equation}
Then, the inner product of two vectors is given by
\begin{align}
	G(u,v) 
	&= G_{\mu\nu} u^\mu v^\nu 
	+ G_{\mu|\kappa\lambda|} u^\mu v_2^{\kappa\lambda} 
	+ G_{|\rho\sigma|\nu} u_2^{\rho\sigma} v^\nu 
	+ G_{|\rho\sigma|\kappa\lambda|} u_2^{\rho\sigma} v_2^{\kappa\lambda} \nonumber\\
	&= \hat{G}_{\mu\nu} \hat{u}^\mu \hat{v}^\nu 
	+ \hat{G}_{\mu|\kappa\lambda|}\hat{u}^\mu \hat{v}_2^{\kappa\lambda} 
	+ \hat{G}_{|\rho\sigma|\nu} \hat{u}_2^{\rho\sigma} \hat{v}^\nu 
	+ \hat{G}_{|\rho\sigma|\kappa\lambda|} \hat{u}_2^{\rho\sigma} \hat{v}_2^{\kappa\lambda}  \, ,
\end{align}
where $\hat{G}$ is a covariant version of the metric $G$. Using the equality between the two lines and eq.~\eqref{eq:CovVec}, its expression in terms of $G$ and the Christoffel symbols can easily be derived. Moreover, since the inner product must be conserved under coordinate transformation, it follows that $\hat{G}_{\mu\nu}$ must transform as a first order (0,2)-tensor, $\hat{G}_{\mu|\kappa\lambda|}$ and $\hat{G}_{|\rho\sigma|\nu}$ as first order (0,3)-tensors and $\hat{G}_{|\rho\sigma|\kappa\lambda|} $ as a first order (0,4)-tensor. In a minimal scenario, where no new fields are introduced, the second order metric can be expressed entirely in terms of the first order metric and its first and second derivatives. Assuming that the connection is metric compatible and torsion-free, this implies that the second order metric is of the following form \cite{Kuipers:2024gfp}
\begin{align}\label{eq:Metric2Order}
	\hat{G}_{\mu\nu} &= g_{\mu\nu} \, ,\nonumber\\
	\hat{G}_{\mu|\kappa\lambda|} &= 0 \, ,\nonumber\\
	\hat{G}_{|\rho\sigma|\nu} &= 0 \, ,\nonumber\\
	\hat{G}_{|\rho\sigma|\kappa\lambda|} &= l_s^{-2} \left[ \frac{1+c_1}{2} g_{\rho\kappa} g_{\sigma\lambda} + \frac{1-c_1}{2} g_{\rho\lambda} g_{\sigma\kappa} - \frac{1-c_2}{n} g_{\rho\sigma} g_{\kappa\lambda} \right]
	+ b_1 \, \mathcal{R}_{\rho\kappa\sigma\lambda} 
	+ b_2 \,  \mathcal{R}_{\rho\lambda\sigma\kappa}
	\nonumber\\
	&\quad
	+ \left( b_3 \, g_{\rho\kappa} g_{\sigma\lambda} + b_4 \, g_{\rho\lambda} g_{\sigma\kappa} + b_5 \, g_{\rho\sigma} g_{\kappa\lambda} \right) \mathcal{R}
	+ b_6 \left( g_{\rho\sigma} \mathcal{R}_{\kappa\lambda} + g_{\kappa\lambda} \mathcal{R}_{\rho\sigma} \right) 
	\nonumber\\
	&\quad
	+ b_7 \left( g_{\rho\lambda} \mathcal{R}_{\sigma\kappa} + g_{\sigma\kappa} \mathcal{R}_{\rho\lambda} \right) 
	+ b_8 \, g_{\rho\kappa} \mathcal{R}_{\sigma\lambda}
	+ b_9 \, g_{\sigma\lambda} \mathcal{R}_{\rho\kappa} \, ,
\end{align}
where $l_s$ is some length scale, $g_{\mu\nu}$ is the first order metric and $\mathcal{R}_{\mu\nu\rho\sigma}$ is the first order Riemann tensor.
Moreover, $b_i,c_i\in\R$ must be fixed by further calculations or constraints. For future purposes, we will also consider the contraction
\begin{equation}\label{eq:MetricContract}
	g^{\rho\sigma} g^{\kappa\lambda} \hat{G}_{|\rho\sigma|\kappa\lambda|}
	= \frac{c_2 \, n}{l_s^2} + c_3 \, \mathcal{R} \, ,
\end{equation}
where 
\begin{equation}
	c_3 = b_1 + b_2 + 2\, b_7 + b_8 + b_9 + n \left(  b_3 + b_4 + 2 \, b_6 \right) +  n^2\, b_5  \, .
\end{equation}

\section{Toy model: scalar particle in arbitrary potential}\label{sec:ToyModel}
In this section, we consider a toy model of a scalar point particle in arbitrary potential described in second order geometry. The main purpose of this section is to discuss how equations of motion arise in second order geometry.

\subsection{Non-relativistic theory}
We consider a theory of a non-relativistic scalar particle in an arbitrary potential on the Riemannian manifold $(\M,g)$. The dynamics is governed by an action
\begin{equation}
	S(x,\mathcal{T}) = \int_{\mathcal{T}} L(x,v,t) \, dt 
\end{equation}
with a classical Lagrangian of the form
\begin{equation}\label{eq:ClassLagNRT}
	L(x,v,t) = \frac{m}{2} g_{ij} v^i v^j + A_i v^i - \mathfrak{U} \, .
\end{equation}
We will now consider the extension of this Lagrangian to second order geometry given by
\begin{equation}\label{eq:LagNRT}
	L(x,\mathfrak{v},\mathfrak{v}_2,t) 
	= 
	L_0(x,\mathfrak{v},\mathfrak{v}_2,t)  + L_\infty(x,\mathfrak{v}_2,t) 
\end{equation}
with
\begin{align}\label{eq:LagNRTFin}
	L_0(x,\mathfrak{v},\mathfrak{v}_2,t) 
	&= 	\frac{m}{2} \, g_{ij} \, \hat{\mathfrak{v}}^i \hat{\mathfrak{v}}^j 
	+ \frac{m}{2} \, \hat{G}_{|ij|kl|} \hat{\mathfrak{v}}_2^{ij} \hat{\mathfrak{v}}_2^{kl}
	+ A_i \hat{\mathfrak{v}}^i 
	+ \frac{1}{2} \hat{\mathfrak{v}}_2^{ij} \nabla_j A_i
	- \mathfrak{U} \, , \\
	L_\infty(x,\mathfrak{v},\mathfrak{v}_2,t) 
	&= \frac{m}{2 \, t} \, g_{ij} \, \hat{\mathfrak v}_2^{ij}\,,\label{eq:LagNRTDiv}
\end{align}
where we introduced the complex velocity field
\begin{equation}
	\mathfrak{v} = v_\circ + \ri \, v_\perp \, .
\end{equation}
Here, we included a logarithmic divergence in the action that can be argued by the following consideration \cite{Nelson:1985,Kuipers:2023pzm}
\begin{align}
	\E \left[ \int \frac{\rd X^i(t)}{dt}  \, \frac{\rd X^j(t)}{dt} \, dt \right]
	&\sim \E \left[ \int \left( v^i \, v^j +  \frac{v_{2}^{ij}}{t} \right) dt \right] ,
\end{align}
and whose relevance to the quantization of the theory will become clear at the end of this section.
\par 

We note that the choice for the extension \eqref{eq:LagNRTFin} is not uniquely defined, and one could consider other second order theories corresponding to the same first order theory \eqref{eq:ClassLagNRT}. However, this particular form \eqref{eq:LagNRTFin} satisfies various nice properties. 
First, it is easy to see that in the classical limit, where $v_\circ\rightarrow v$, $v_\perp\rightarrow 0$ and $v_{2}\rightarrow 0$, one obtains the classical Lagrangian \eqref{eq:ClassLagNRT}. 
Second, the Lagrangian is manifestly covariant. Third, under time-reversal symmetry, the roles of the forward and backward derivatives exchange:
\begin{equation}
	v_+ \leftrightarrow v_- \qquad \Rightarrow\qquad  v_\circ \leftrightarrow v_\circ , \quad v_\perp \leftrightarrow - v_\perp
\end{equation}
such that
\begin{equation}
	L(x,\mathfrak{v},\mathfrak{v}_2,t)  \overset{{\rm time-reversal}}{\longrightarrow} L(x,\bar{\mathfrak{v}},\bar{\mathfrak{v}}_2,t) = \bar{L}(x,\mathfrak{v},\mathfrak{v}_2,t) \, .
\end{equation}
Finally, the theory is invariant under gauge-transformations. Indeed, the addition of a total derivative term to the action can be compensated by a redefinition of the potentials:
\begin{align}
	S 
	&\rightarrow S + \int_{\mathcal{T}} \rd_2\chi(x,t)
	\nonumber\\
	&=
	\int_{\mathcal{T}} L(x,v,t) \, dt 
	+ \int_{\mathcal{T}}  \left(\p_t \chi + \mathfrak{v}^i \p_i \chi + \frac{1}{2} \mathfrak{v}_2^{ij} \p_j \p_i \chi \right) dt \nonumber\\
	&= 
	\int_{\mathcal{T}} \tilde{L}(x,v,t) \, dt \, ,
\end{align}
where the Lagrangian $\tilde{L}$ depends on the potentials
\begin{equation}\label{eq:GaugeTrans}
	\tilde{A}_i = A_i + \p_i \chi  
	\qquad {\rm and} \qquad
	\tilde{\mathfrak{U}} = \mathfrak{U} - \p_t \chi \, .
\end{equation}
\par 

In order to obtain the dynamics of the theory, one must derive the equations of motion from this action. In the Hamilton-Jacobi formalism, one can define Hamilton's principal function by
\begin{equation}\label{eq:HamPrinc}
	S_{H}(x,t;x_0,t_0) = \E \left[ \int_{t_0}^t L[X(s),\mathfrak{v}(X(s),s),\mathfrak{v}_2(X(s),s),s] \, ds \; \Big| \; X(t)=x,\, X(t_0)=x_0 \right] .
\end{equation}
We emphasize that due to the presence of the divergent term in the Lagrangian, this function is not single valued. Indeed, using the split \eqref{eq:LagNRT}, one finds
\begin{equation}\label{eq:HamPrincSplit}
	S_{H}(x,t;x_0,t_0) = 
	S_{H0} + S_{H\infty}\, ,
\end{equation}
where $S_{H0}$ is a single valued function, but the integrand $S_{H\infty}$ contains a pole, generating a quantization that is associated to the winding around this pole. Clearly, in first order geometry, the second term vanishes and no such quantization occurs. 
\par

The Hamilton-Jacobi equations can easily be derived and are given by \cite{Kuipers:2023pzm}
\begin{align}
	\frac{\p S_H}{\p x^i} &= \hat{\mathfrak{p}}_i \, ,\\
	\frac{\p S_H}{\p t} &= L - \hat{\mathfrak{p}}_i \hat{\mathfrak{v}}^i - \frac{1}{2} \, \hat{\mathfrak{v}}_2^{ij} \nabla_j \hat{\mathfrak{p}}_i\, ,
\end{align}
where the momentum is defined as
\begin{equation}
	\hat{\mathfrak{p}}_i = \frac{\p L}{\p \hat{\mathfrak{v}}^i} \, .
\end{equation}
\par

We may now apply the equations of motion to the Lagrangian \eqref{eq:LagNRT}. In particular, we apply the covariant derivative to the second Hamilton-Jacobi equation and plug in the first Hamilton-Jacobi equation and the Lagrangian \eqref{eq:LagNRT}. Assuming that the connection is metric compatible and torsion free, this yields
\begin{align}\label{eq:EomNRT}
	\left\{ 
		m \, g_{ij} \left[
			\p_t  
			+ \hat{\mathfrak{v}}^k \nabla_k 
			+ \frac{1}{2} \hat{\mathfrak{v}}_2^{kl}  \nabla_l \nabla_k 
		\right] 
		+ m \,\p_t(g_{ij})
		+ \frac{m}{2} \left[
			g_{jk} \nabla_i (\hat{\mathfrak{v}}_2^{kl}) \nabla_l 
			+ \hat{\mathfrak{v}}_2^{kl} \mathcal{R}_{ilkj} 
		\right]
		- F_{ij} 
	\right\} \hat{\mathfrak{v}}^j
	\nonumber\\
	=
	\frac{1}{2} \, \hat{\mathfrak{v}}_2^{jk} \nabla_k F_{ij} 
	- \p_t A_i 
	-\nabla_i \mathfrak{U}
	+ \frac{m}{2} \nabla_i \left( \hat{G}_{|jk|lm|} \hat{\mathfrak{v}}_2^{jk} \hat{\mathfrak{v}}_2^{lm}\right)
	+ \frac{m}{2 \, t} g_{jk} \nabla_i \hat{\mathfrak{v}}_2^{jk} \, ,
\end{align}
where $F_{ij} = \nabla_i A_j - \nabla_j A_i$ is the field strength.
\par 

After fixing the second order velocity field $\hat{\mathfrak{v}}_{2}$, one can in principle solve the equation of motion \eqref{eq:EomNRT} for the velocity field $\hat{\mathfrak{v}}$.  For any such solution one can then obtain a solution for $X(t)$ by solving the differential equations \eqref{eq:StratEq} with $v_\circ = {\rm Re}\left( \hat{\mathfrak{v}}^i - \frac{1}{2} \Gamma^i_{jk}\hat{\mathfrak{v}}_2^{jk}\right)$ and $v_\perp= {\rm Im}\left( \hat{\mathfrak{v}}^i - \frac{1}{2} \Gamma^i_{jk}\hat{\mathfrak{v}}_2^{jk}\right)$.
\par 

Let us now fix
\begin{equation}
	\hat{\mathfrak{v}}_{2}^{ij} = - \frac{\alpha}{m} \, g^{ij} \, .
\end{equation}
with $\alpha\in\mathbb{C}$ and define a wave-function by
\begin{equation}
	\Psi(x,t) = \exp\left[ - \frac{S_H}{\alpha} + \frac{c_2 \, \alpha\,  n}{2\,m\,l_s^2}\, t \right] .
\end{equation}
Then, using eq.~\eqref{eq:MetricContract}, one finds that the Hamilton-Jacobi equations are equivalent to the complex diffusion equation
\begin{equation}\label{eq:Diffusion}
	\alpha \, \frac{\p \Psi}{\p t} 
	= \left[ 
		\frac{g^{ij}}{2\,m} 
		\left( \alpha \nabla_i + A_i\right) 
		\left( \alpha \nabla_j + A_j\right) 
		- \frac{c_3 \, \alpha^2}{2\,m} \, \mathcal{R} 
		+ \mathfrak{U} 
	\right] \Psi \, .
\end{equation}
It immediately follows that the gauge transformation \eqref{eq:GaugeTrans} is related to the local symmetry transformation of the wave function
\begin{equation}
	\Psi\rightarrow \tilde{\Psi} = \Psi \,e^{\chi/\alpha} \, .
\end{equation}
Finally, using the condition \eqref{eq:V2fix} and the split \eqref{eq:HamPrincSplit}, the wave function is given by 
\begin{equation}
	\Psi(x,t) = \exp\left[ - \frac{S_{H0}}{\alpha} + \pi \, \ri \, \sum_{i=1}^n k_i + \frac{\alpha\, c_2 \, n}{2\,m\,l_s^2}\, t \right]
	\qquad 
	k_i \in\mathbb{Z} \, , 
\end{equation}
It follows that only by including the divergent term \eqref{eq:LagNRTDiv}, one obtains an equivalence between the solutions (up to a $\pm$ sign) of the diffusion equation \eqref{eq:Diffusion} and solutions of the system \eqref{eq:EomNRT}. If this term is not included, Hamilton's principal function provides only one of the branches of $\ln\Psi$. Moreover, since the branches of $\ln\Psi$ relate to the energy levels, it follows that the quantization of Hamilton's principal function is related to the energy eigenstates of this complex diffusion equation.

\subsubsection{Geodesics}\label{sec:Geodesics}
The toy model yields the equation of motion given by eq.~\eqref{eq:EomNRT}. Let us now discuss this equation in some detail. For simplicity, we will consider the case of vanishing potentials, i.e. $A_i=\mathfrak{U}=0$, a metric as given in eq.~\eqref{eq:Metric2Order} and set, as in eq.~\eqref{eq:SecVel},
\begin{equation}\label{eq:V2fix}
	\hat{\mathfrak{v}}_2^{ij}=\frac{|\alpha|e^{\ri\phi}}{m} g^{ij} \, .
\end{equation}
Then, eq.~\eqref{eq:EomNRT} becomes
\begin{equation}\label{eq:EomNRT2}
	\left[ 
		g_{ij}
		\begin{pmatrix}
			\p_t + \hat{v}_\circ^k \nabla_k
			& - \hat{v}_\perp^k \nabla_k \\
			\hat{v}_\perp^k \nabla_k & \p_t + \hat{v}_\circ^k \nabla_k
		\end{pmatrix}
		+ 
		\frac{|\alpha|}{2m}
		\begin{pmatrix}
			\cos\phi & -\sin\phi \\
			\sin \phi & \cos\phi
		\end{pmatrix}
		\left( g_{ij} \Box - \mathcal{R}_{ij} \right)
	\right]
	\begin{pmatrix}
		\hat{v}_\circ^j \\ \hat{v}_\perp^j
	\end{pmatrix}
	=
	\frac{|\alpha|c_3}{2m} 
	\begin{pmatrix} 
		\cos\phi^2 - \sin \phi^2 \\ 2 \cos\phi \sin\phi
	\end{pmatrix} 
	\nabla_i \mathcal{R} \, .
\end{equation}
By taking the first order limit $(\hat{\mathfrak{v}},\hat{\mathfrak{v}}_2)\rightarrow (v,0)$, one obtains the geodesic equation from first order geometry:
\begin{align}
	\frac{D v^i}{dt} 
	= \left( \p_t  + v^j \nabla_j \right) v^i
	= \frac{d v^i}{dt} + \Gamma^i_{jk} v^j v^k
	=	0 \, .
\end{align}
Therefore, eq.~\eqref{eq:EomNRT2} can be interpreted as the second order generalization of the geodesic equation. 
\par

We will now make a few observations. First, we find that the directional derivative is modified, such that it contains second order derivatives. Due to the presence of these second order derivatives, the geodesics incorporate a wave-like behavior, as shown by the equivalence with the diffusion equation \eqref{eq:Diffusion}.
Second, we find that the curvature acts as a potential for the second order geodesic equation, which corresponds to a non-minimal coupling of the particle to the gravitational field in the diffusion equation \eqref{eq:Diffusion}.
\par 

Third, we find that the geodesic equation does not only provide an equation for the vector field $v_\circ$, but also for the vector field $v_\perp$. The two vector fields are defined in section \ref{sec:Rough}, but can be reinterpreted using the worldsheet property. We may thus think of the set of paths as a 2 dimensional field $X(t,s)$ depending on two parameters, and interpret the velocities as 
\begin{align}
	v_\circ \sim \frac{\p X}{\p t} \qquad {\rm and} \qquad v_\perp \sim \frac{\p X}{\p s} \, .
\end{align}
Following this reinterpretation, one finds that the current velocity $v_\circ$ denotes the velocity along the paths in the path integral, whereas the osmotic velocity $v_\perp$ measures the relative velocity between the paths on this worldsheet. Within this intepretation, the kinetic term in the Lagrangian can be rewritten by introducing a metric on this worldsheet. Therefore, since the shape of eq.~\eqref{eq:EomNRT2} depends on the choice of Lagrangian \eqref{eq:LagNRTFin}, one may reinterpret this as a dependence on the worldsheet metric.

\subsection{Relativistic theory}\label{sec:RLT}
The relativistic theory can be obtained from the non-relativistic theory by replacing the time parameter $t$ by an arbitrary affine parameter $\lambda$, the space coordinates $\{x^i\,|\,i\in\{1,2,...,n\}\}$ by spacetime coordinates $\{x^\mu\,|\,\mu\in\{0,1,...,n-1\}\}$ and imposing invariance under affine reparameterizations. In practice, this means that the classical Lagrangian \eqref{eq:ClassLagNRT} is replaced by the relativistic Lagrangian \cite{Lust:1989tj}
\begin{equation}
	L(x,v,\varepsilon) = \frac{1}{2\, \varepsilon} \, g_{\mu\nu} v^\mu v^\nu + A_\mu  v^\mu  - \frac{\varepsilon\, m^2}{2} \, ,
\end{equation}
where the field $\varepsilon$ can be gauge fixed in the equations of motion.
The relativistic second order Lagrangian corresponding to \eqref{eq:LagNRT} is then given by
\begin{equation}
	L(x,\mathfrak{v},\mathfrak{v}_2,\varepsilon) 
	= 
	\frac{g_{\mu\nu}}{2\, \varepsilon} \, \hat{\mathfrak{v}}^\mu \hat{\mathfrak{v}}^\nu 
	+ \frac{g_{\mu\nu}}{2\, \varepsilon\, \lambda} \, \hat{\mathfrak{v}}_2^{\mu\nu} 
	+ \frac{\hat{G}_{|\mu\nu|\rho\sigma|}}{2\, \varepsilon} \, \hat{\mathfrak{v}}_2^{\mu\nu} \hat{\mathfrak{v}}_2^{\rho\sigma}
	+ A_\mu \hat{\mathfrak{v}}_2^\mu 
	+ \frac{1}{2} \hat{\mathfrak{v}}_2^{\mu} \nabla_\nu A_\mu 
	- \frac{\varepsilon\, m^2}{2}  \, ,
\end{equation}
and the corresponding equation of motion \eqref{eq:EomNRT} becomes
\begin{align}\label{eq:EomRLT}
	\left\{ 
		g_{\mu\nu} \left[ 
			\hat{\mathfrak{v}}^\rho \nabla_\rho + \frac{1}{2} \hat{\mathfrak{v}}_2^{\rho\sigma}  \nabla_\sigma \nabla_\rho 
		\right] 
		+ \frac{1}{2} \left[
			g_{\nu\rho} \nabla_\mu (\hat{\mathfrak{v}}_2^{\rho\sigma}) \nabla_\sigma 
			+ \hat{\mathfrak{v}}_2^{\rho\sigma} \mathcal{R}_{\mu\sigma\rho\nu} 
		\right]
		- \varepsilon \, F_{\mu\nu} 
	\right\} \hat{\mathfrak{v}}^\nu
	\nonumber\\
	=
	\frac{\varepsilon}{2} \, \hat{\mathfrak{v}}_2^{\nu\rho} \nabla_\rho F_{\mu\nu}
	+ \frac{1}{2} \, \nabla_\mu \left( 
		\hat{G}_{|\nu\rho|\sigma\kappa|} \hat{\mathfrak{v}}_2^{\nu\rho} \hat{\mathfrak{v}}_2^{\sigma\kappa} 
	\right) 
	+ \frac{g_{\nu\rho}}{2\, \varepsilon \, \lambda} \nabla_\mu \hat{\mathfrak{v}}_2^{\nu\rho}. 
\end{align}
\par

As for the non-relativistic theory, one may fix
\begin{equation}
	\hat{\mathfrak{v}}_{2}^{\mu\nu} = - \alpha \, \varepsilon \, g^{\mu\nu} \, .
\end{equation}
and define a wave-function by
\begin{align}
	\Psi(x,\tau) &=  \Phi(x) \, \exp\left[ - \frac{\tau}{2} \left( \frac{m}{\alpha} - \frac{\alpha\, c_2 \, n}{m\,l_s^2} \right) \right]\, ,\\
	\Phi(x) &= \exp\left( - \frac{S}{\alpha} \right)
\end{align}
with $\alpha\in\mathbb{C}$, $x$ the spacetime coordinate and $\tau$ the proper time.
Then a straightforward calculation shows that that the Hamilton-Jacobi equations are equivalent to the complex wave equation
\begin{equation}\label{eq:Wave}
	\frac{\varepsilon}{2} \left[ 
		g^{\mu\nu} \left( \alpha \nabla_\mu + A_\mu\right)  \left( \alpha \nabla_\nu + A_\nu \right) 
		- c_3 \, \alpha^2 \, \mathcal{R} 
		+ m^2
	\right] \Phi
	=
	0 \, .
\end{equation}
Using the condition \eqref{eq:V2fix} and the split \eqref{eq:HamPrincSplit} the wave function can be rewritten as
\begin{equation}\label{eq:Wave}
	\Phi(x) = \exp\left[ - \frac{S_{H0}}{\alpha} + \pi \, \ri \, \sum_{\mu=0}^n k_\mu \right]
	\qquad 
	k_\mu \in\mathbb{Z} \, .
\end{equation}
By the same reasoning as in the non-relativistic theory, it follows that the quantization of Hamilton's principal function is related to various branches of $\ln\Phi$, which, in the relativistic theory, can be related to the particle number eigenstates of this complex wave equation.

\subsubsection{Energy-momentum relation}
In the relativistic theory, the equation of motion for the auxiliary field $\varepsilon$ provides a constraint given by
\begin{equation}
	g_{\mu\nu} \, \hat{\mathfrak{v}}^\mu \hat{\mathfrak{v}}^\nu  + \hat{G}_{|\mu\nu|\rho\sigma|} \hat{\mathfrak{v}}_2^{\mu\nu} \hat{\mathfrak{v}}_2^{\rho\sigma}
	=
	- \varepsilon^2 \, m^2 \, .
\end{equation}
This constraint can be rewritten as an energy-momentum relation:
\begin{equation}
	g^{\mu\nu} \left( \hat{\mathfrak{p}}_\mu - A_\mu \right) \left( \hat{\mathfrak{p}}_\nu - A_\nu \right) + \varepsilon^{-2}\hat{G}_{|\mu\nu|\rho\sigma|} \hat{\mathfrak{v}}_2^{\mu\nu} \hat{\mathfrak{v}}_2^{\rho\sigma}
	= - m^2
\end{equation}
Since $\mathfrak{p}=p + \ri \, q$ is a complex momentum, this reduces to two conditions. Thus, after setting $A=0$ and fixing the second order velocity as in eq.~\eqref{eq:V2fix} and using the expression for the second order metric \eqref{eq:MetricContract}, we obtain
\begin{align}
	g^{\mu\nu} \left( \hat{p}_\mu \hat{p}_\nu - \hat{q}_\mu \hat{q}_\nu \right) + |\alpha|^2 \left(\cos(\phi)^2 - \sin(\phi)^2 \right) \left( c_2 \, n \, l_s^{-2} + c_3 \, \mathcal{R} \right)
	&= - m^2 \, ,\\
	g^{\mu\nu} \hat{p}_\mu \hat{q}_\nu + |\alpha|^2 \cos(\phi) \sin(\phi) \left( c_2 \, n \, l_s^{-2} + c_3 \, \mathcal{R} \right)
	&= 0 \, .
\end{align}
We note that this relation suggests a deformation of the Lorentz symmetry in second order geometry, such that the usual first order Lorentz group is extended to a second order Lorentz group. Given the strong constraints on Lorentz symmetry violations, one expects that this second order Lorentz symmetry is broken at some energy scale, such that below this scale the usual Lorentz symmetry is recovered. In that scenario, the first relation splits into the first order energy-momentum relation and an additional constraint:
\begin{align}
	g^{\mu\nu} \hat{p}_\mu \hat{p}_\nu
	&= m^2 \, ,\\
	g^{\mu\nu} \hat{q}_\mu \hat{q}_\nu &= |\alpha|^2 \left(\cos(\phi)^2 - \sin(\phi)^2 \right) \left( c_2 \, n \, l_s^{-2} + c_3 \, \mathcal{R} \right) .
\end{align}
The second equation provides an energy-momentum relation for the momenta $q$ with a mass term that depends on the curvature $\mathcal{R}$ and the second order data $\alpha$ and $l_s$.

\section{Conclusion}
In this work, we reviewed the motivation for extending differential geometry to higher order for studying the interplay between quantum and/or statistical theories and gravity. Then, we reviewed the basic mathematical ingredients of such a higher order geometry and discussed how second order geometry can be employed in physics by studying a toy model.
In this toy model, discussed in section \ref{sec:ToyModel}, we found that the classical equations of motion are modified by the inclusion of terms second order in derivative, cf. eqs. \eqref{eq:EomNRT} and \eqref{eq:EomRLT}. Therefore, the geodesic equations in second order geometry naturally incorporate the wave-like behavior of quantum theories as discussed in section \ref{sec:Geodesics}.
\par 

As shown by the toy model, the Lagrangian is extended in second order geometry by the introduction of second order velocity fields. This may seem paradoxical, as this extension will modify the path integral measure \eqref{eq:PImeasure}, whereas the path integral formulation served as the primary motivation for developing second order geometry. However, this does not necessarily lead to problematic features \cite{Kuipers:2024gfp}. Indeed, on a flat spacetime, the first and second order sector decouple, such that the second order part of the Lagrangian only affects the normalization of the path integral, whereas the dynamics is still described by the first order part. Therefore, on a flat spacetime one expects that a path integral measure based on the second order Lagrangian yields the same correlation functions as one based on a first order Lagrangian. In curved spacetime, on the other hand, the first and second order sector couple. Consequently, from the second order perspective, a  path integral measure based on a first order Lagrangian does not respect general covariance and will generate gravitational anomalies. General covariance can be restored by employing the second order Lagrangian in the path integral measure, which may allow to avoid such anomalies.

\section{Outlook}
We will split the outlook section into two parts. In the first part, we discuss further research directions specified to higher order geometry and its application in studying the interplay between gravity and quantum theory. In the second part \ref{OutlookII}, we will provide an overview of some relations with other theoretical frameworks that have been used to study this interplay.

\subsection{Second and higher order geometry}

\begin{itemize}
	\item \textbf{Field theories}: In section \ref{sec:ToyModel}, we discussed a toy model of a scalar point particle in second order geometry, and saw how the equations of motion are modified in this case. This worldline theory of a point particle may also be regarded as a field theory in one dimension. An important open question is how this 1-dimensional field theory in second order geometry should be extended to field theories in dimension $n\geq2$. One observation one can make in this regard is that the configuration space will change compared to first order geometry. The configuration space of classical field theories is the first order jet bundle, such that the Lagrangian $L(\phi,\nabla\phi)$ is a function of the fields and their first order derivatives only. In second order geometry, on the other hand, two fields are no longer in the same equivalence class if their second derivative differs, due to the modification of the total derivative \eqref{eq:TotDer}. Therefore, the configuration space may be given by the second order jet bundle \cite{Huang:2022}, such that a Lagrangian $L(\phi,\nabla\phi,\nabla\nabla\phi)$ depends on the field and both the first and second order derivatives. More generally, for $k$th order geometry, one expects the $k$th order jet bundle to arise as the configuration space. Classical field theories have been studied in this context, for example in \cite{Campos:2009ue,Campos:2010ay}.
	\item \textbf{Spin}: In section \ref{sec:ToyModel}, we studied a toy model for a scalar theory. An important extension of this theory would be the study of a worldline theory of a particle with spin. For such an extension, one expects that the wave equation \eqref{eq:Wave} will be replaced by a Dirac equation for spin-1/2. Moreover, given the fact that Poisson process can be decomposed into spin operators, as discussed in section \ref{sec:StochCalc}, one expects that this requires the study of discrete theories, which suggests an extension to infinite order geometry.
	From the geometrical perspective such studies will require a study of spinor bundles in higher order geometry.
	\item \textbf{Lorentz symmetry}: As discussed in section \ref{sec:2ndorder}, the structure group of the tangent bundle is extended in second order geometry from the general linear group to the It\^o group. Similarly, the group of transformations that keep lengths invariant will be extended in second order geometry. In section \ref{sec:RLT}, we saw how this is reflected by an extended energy-momentum relation. In future research, it would be interesting to study the structure of this second order Lorentz group in more detail. Additionally, it would be interesting to study the possibility of breaking this second order Lorentz symmetry down to the first order Lorentz symmetry.
	\item \textbf{Lie derivatives}: The Lie bracket of a vector of order $k$ with a vector of order $l$ yields a vector of order $(k+l-1)$. Therefore, the Lie bracket closes upon itself within first order geometry, but not in higher order geometry. For certain subclasses of tensors, one can still define a Lie derivative. For example, the Lie derivative of a $1$st order vector along a with a $k$th order vector yields a $k$th order vector \cite{Kuipers:2021jlh,Kuipers:2022wpy,Huang:2022}. However, if one wants to define a Lie bracket for any vector of order $k$ with a vector of order $l$, one must either do this within infinite order geometry \cite{Bies:2023zvs} or modify the Lie bracket, such that it closes upon itself. 
	\item \textbf{Geodesic deviation} As shown in section \ref{sec:ToyModel}, the geodesic equations are modified by the incorporation of second order derivatives. Following up on this, it would be interesting to study the modification of geodesic deviation equations in second order geometry. Such a study could also serve as a first step towards studying the second order modification of the Raychaudhuri equations, which could then provide insight in the modification of singularity theorems within second order geometry.
\end{itemize}

\subsection{Relations to other frameworks}\label{OutlookII}
Various aspects of second and higher order geometry have counterparts in other theoretical frameworks that have been applied in the study of the interplay between quantum theory and gravity. Here, we present a non-exhaustive list of some of these frameworks. The discussion will be superficial but may help in understanding the relation of higher order geometry with these other frameworks, and could serve as a starting point in the study of common features of higher order geometry and other frameworks.
\begin{itemize}
	\item \textbf{String theory}: The worldsheet property provides a natural relation to string theory: since the worldlines are replaced by rough worldsheets, one may think of these worldsheets as generated by a propagating string. We may push this analogy slightly further using the stringy reinterpretation of the velocity fields $v_\circ,v_\perp$ in section \ref{sec:Geodesics}. However, beyond these elementary considerations, the two frameworks deviate: in string theory, one would assume differentiability of a classical worldsheet, define a Polyakov action $S=\int\int L(x,v_\circ,v_\perp)\, d\tau d\sigma$, and quantize the theory. In the toy model from section \ref{sec:ToyModel}, on the other hand, the worldsheet arises due to the path integral quantization of the worldline theory. As the theory is already quantized, the worldsheet has a rough structure, and its dynamics is governed by an action $S=\E[\int_\mathcal{T} L \, d\tau]= \int_\Omega\int_\mathcal{T} L \, d\tau \rd\mathbb{P}(\omega)$ that averages over the paths in the path integral.
	\item \textbf{Generalized geometry}: As discussed in section \ref{sec:2ndorder}, $k$th-order geometry studies an extended tangent bundle structure of the form $\tilde{T}\M = \bigsqcup_{x\in\M}\bigoplus_{l=1}^k T^{l}(T_x\M)$. There exist other geometrical frameworks where extended tangent bundle structures are studied. An example is generalized geometry  \cite{Hitchin:2003cxu,Gualtieri:2003dx,Gualtieri:2007bq}, where one considers the tangent bundle $\tilde{T}\M = T_x\M\oplus T_x^\ast\M$. Clearly, these two extensions are different, such that the only similarity between the frameworks seems to be that they build a geometry on some extension of the tangent spaces. However, as discussed in section \ref{sec:2ndorder}, higher order geometry also increases the number of velocity fields due to the fact that the left and right limit do not coincide. Consequently, in second order geometry, the tangent spaces can be extended to eq.~\eqref{eq:Tangentspace}. Focusing on the first part of this space, we find a tangent space $\tilde{T}_x\M=T_x\M\oplus T_x\M$, with corresponding velocities $(v_+,v_-)$ or equivalently $(v_\circ,v_\perp)$. Then, using the metric one may map $g^\flat(v_\perp)\in T^\ast_x\M$, which suggests that generalized geometry may be incorporated in the second order tangent bundles that were considered in this work.
	Generalized geometry has also obtained applications in string theory, as it can be embedded in double field theory \cite{Siegel:1993th,Hull:2009mi,Hull:2009zb,Hohm:2010jy,Hohm:2010pp,Hohm:2013bwa} by extending the dimension of the manifold such that it matches the dimension of the extended tangent space.
	\item \textbf{Non-commutative geometry}: Both higher order geometry and non-commutative geometry, e.g. \cite{Beggs:2020}, propose that one must extend the notion of geometry beyond  differential geometry. The particular type of modification is very different in the two frameworks, but some relations can be found. Indeed, in infinite order geometry, trajectories become discontinuous and a relation with the Lie-algebra valued type of spacetime non-commutativity with commutators of the form $[x^\mu,x^\nu]= C^{\mu\nu}_{\rho} x^\rho$ can be established \cite{Biane:2010,Arzano:2024apl}. The Moyal type of non-commutativity with commutators of the form $[x^\mu,x^\nu]= B^{\mu\nu}$ does not seem to arise in this infinite order limit. However, the latter type may be implemented within second order geometry by allowing for non-symmetric vector fields $v_2^{\mu\nu}$ \cite{Kuipers:2024gfp}.
	\item \textbf{Extended tangent bundles and unification of forces}: One of the main ingredients of higher order geometry is that $\dim (T_x\M)>\dim (\M)$. Thus, within the higher order geometry paradigm, spacetime may remain $3+1$ dimensional, but the tangent spaces must have larger dimension. The observation that there may be a discrepency between the dimensions of the manifold and the tangent space has been made independently of higher order geometry, and has been applied to the study of a unification of forces \cite{Percacci:1984ai,Percacci:1990wy,Nesti:2009kk,Chamseddine:2010rv,Chamseddine:2013hwa,Chamseddine:2016pkx,Krasnov:2017epi,Konitopoulos:2023wst,Roumelioti:2024lvn}.
	Also, within the context of higher order geometry such a unification has been studied \cite{Bies:202406,Bies:2024gjn}.
	\item \textbf{Area metrics}: The second order part of the metric $\hat{G}_{|\rho\sigma|\kappa\lambda|}$ constructed in eq.~\eqref{eq:Metric2Order} is an area metric for certain values of $b_i,c_i$. For example, for $\hat{G}_{|\rho\sigma|\kappa\lambda|}$ to be an area metric the constraints on $c_i$ are $c_1=\pm1$ and $c_2=1-n$. Therefore, the geometry constructed from the second order part of the metric only can be related to that of area manifolds, which has found various applications in quantum gravity theories \cite{Schuller:2005yt,Schuller:2005ru,Punzi:2006hy,Punzi:2006nx,Dittrich:2023ava,Borissova:2023yxs}.
	\item \textbf{Entropic gravity}: In the entropic gravity theory based on a relative entropy principle \cite{Bianconi:2024aju,Bianconi:2025rnd}, one constructs an extended metric field, which is similar to the one considered in section \ref{sec:Metric}. More precisely, in that case one considers a metric of the form $\tilde{G}= G \oplus (G_{\mu\nu} \rd x^\mu \otimes \rd x^\nu) \oplus [G_{\mu\nu\rho\sigma} (\rd x^\mu \wedge \rd x^\nu ) \otimes (\rd x^\rho \wedge \rd x^\sigma)]$. Hence, as in second order geometry it contains both a first order part $G_{\mu\nu}$ and a second order part $G_{\mu\nu\rho\sigma}$. However, in entropic gravity the second order part is an area metric, while it is not clear whether this must be true in second order geometry. Moreover, in entropic gravity the metric contains a scalar part $G$, which does not have a clear counterpart in second order geometry.
	\item \textbf{Quadratic Gravity}: A second order gravitational theory will depend on the entire metric field $\hat{G}$, thus both on the first order part $g_{\mu\nu}$ and on the second order part $\hat{G}_{|\rho\sigma|\kappa\lambda|}$. As the latter part contains second order derivatives of the metric, the resulting Lagrangian will contain terms that are linear and terms that are quadratic in the curvature tensor, similar to the theory of quadratic gravity \cite{Stelle:1976gc,Stelle:1977ry}. This is a renormalizable theory of gravity, but it contains an Ostragradski instability, due to the appearance of higher derivative terms. However, the proof for this instability relies on the integration by parts formula from first order geometry. Since the integration by parts formula is generalized in higher order geometry, higher derivative theories may be free from such an instability when defined within the context of higher order geometry. In fact, the worldline theory, discussed in section \ref{sec:ToyModel}, does not suffer from such an instability. This suggests that the gravitational theory in second order geometry may be both renormalizable and ghost-free. We note that similar arguments for the absence of ghosts have been given in entropic gravity \cite{Bianconi:2024aju} and postquantum gravity \cite{Oppenheim:2018igd,Grudka:2024llq}.
\end{itemize}

\section*{Acknowledgments}
It is a pleasure to thank Ramy Brustein and Dieter L\"ust for useful discussions following the presentation of this work at the Corfu workshop. Furthermore, I am grateful to the Alexander von Humboldt foundation for the financial support.

\end{document}